\documentclass[aps,prl,twocolumn,showpacs]{revtex4}
\usepackage{graphicx}
\begin{document}

\title{Domain instability during precessional magnetization reversal}

\author{A. Kashuba}

\affiliation{L. D. Landau Institute for Theoretical Physics, Russian Academy of Sciences, 2 Kosygina str., 119334 Moscow}

\date{\today}

\begin{abstract}

Spin wave equations in the non-equilibrium precessing state of a ferromagnetic system are found. They show a spin-wave instability towards growing domains of stable magnetization. Precession of the uniform magnetization mode is described by the Landau Lifshitz equation with the exponentially growing in time effective Gilbert dissipation constant that could have both signs. On the developed stages of the domain instability a non-stationary picture of domain chaos is observed.

\end{abstract}


\maketitle

Recently fast pulses of strong magnetic field have brought a possibility to study experimentally a precessional magnetization reversal \cite{back:98,back:99,tudosa:04}. The magnetic field of such a pulse is spatially uniform on the lengths that are typical for a ferromagnetic system: the domain wall thickness, the grain size or the width of thin ferromagnetic films etc. The amplitude of the magnetic field in the pulse is such that the Zeeman energy exceeds the anisotropy energy that is present in the ferromagnetic system. The integral of the magnetic field over the duration of the pulse could be varied to result in a uniform rotation of spins by angles ranging from few degrees to up to $2\pi$ \cite{back:98}. There are different possibilities how to apply the magnetic field vector with respect to the anisotropy axis of a ferromagnetic system \cite{back:99}. Also little change occurs to the ferromagnetic order of rotating spins during the pulse. All these special properties of the fast pulses allows one to prepare in effect a non-equilibrium uniform ferromagnetic state with the magnetization vector pointing in an arbitrary direction. An alternative method to reverse magnetization is an application of longer pulses of weaker magnetic field where magnetization precession results from the combined action of the magnetic field and the anisotropy during the pulse \cite{silva:99,gerrits:02,freeman:02,kaka:02,schumacher:03}.

In general after the pulse ceases a precession of uniform magnetization will start due to the anisotropy with the magnetization unit vector sweeping large angles. There is an emerging experimental evidence that such a precession follows the Landau Lifshitz equation with the Gilbert dissipation constant exceeding by tens times the Gilbert dissipation constant in the same ferromagnetic system under the conditions of the Ferro-Magnetic Resonance \cite{stamm:05} where small angles of precession are excited. There are two special non-equilibrium ferromagnetic states. First, in some systems the dissipation is especially weak and the large angle precession is periodic in time. Second, the magnetization unit vector after the pulse is rotated close to the maximum or saddle point of the anisotropy energy. Obviously, in the last case a fundamental instability towards growing domains of stable magnetization arises with the domain sizes being of the order of the domain wall thickness. The initial stages of this domain instability can be described in terms of the growing spin waves. In the first case in general a spin wave instability also develops like the two-magnon Suhl instability that has been extensively studied in connection with the parametric excitation of spin waves in a ferromagnet in equilibrium \cite{white}. Recently such instability has been shown to be responsible for a large effective damping during periodic precession in the area of angles around the stability point \cite{dobin:03}. 

In this letter I find a system of spin wave equations in the rotating frame that follows the uniform mode precession of the non-equilibrium state of ferromagnetic system. Solution of these equations under certain conditions yields a spin-wave instability.  The equation for uniform mode precession coupled to the dynamics of growing instable spin waves is the Landau-Lifshitz equations with effective Gilbert constant that exponentially grows in time too. On the developed stages of instability ab-initio numerical solution of the Larmor precession in a cluster of exchange coupled classical spins shows a non-stationary picture of domain chaos. 

Consider a ferromagnetic particle in a precessional state that is described by the uniform time-dependent ferromagnetic order parameter unit vector $\vec{n}(t)$. Spin waves are also present in this particle either due to the thermal excitation at temperature $T$ before arrival of the pulse or due to the spin wave instabilities developing after the pulse. The local ferromagnetic order parameter unit vector $\vec{M}(t,\vec{r})= \vec{n}(t) \sqrt{1-\phi^a\phi_a}+ \phi^a\vec{e}_a(t)$ ($\vec{M}^2=1$) is parameterized by spin wave two-component field $\phi^a(t,\vec{r})$ with $a=1,2$ and by a rotating frame of three mutually orthogonal vectors $\vec{n}$, $\vec{e}_a$ at any moment of time. A specific choice of two vectors $\vec{e}_a(t)$ is dictated by the vector $\vec{n}(t)$ apart from the arbitrary time-dependent rotation about the axis $\vec{n}$. This rotation represents $U(1)$ gauge invariance in the description of spin waves in a ferromagnet. It is convenient to relate the rotating frame to the curve $\vec{w}(t)$ in the spin space. The vector $\vec{n}(t)$ is the unit tangent vector to this curve: $d\vec{w}/dt= \vec{n}(t)$, the vector $\vec{e}_1(t)$ being the unit principal normal vector to this curve and $\vec{e}_2(t)$ being the unit principal binormal vector to this curve. This choice imposes a gauge fixing condition known as the Frenet-Serret equations: $d\vec{n}/dt=\vec{\zeta}\times \vec{n}$, $d\vec{e}_a/dt= \vec{\zeta}\times \vec{e}_a$, where the rotating frame is precessing as a solid structure about the Darboux vector $\vec{\zeta}(t)= -\kappa\vec{n}+k\vec{e}_2$, where $k(t)$ is the local curvature of the curve and $\kappa(t)$ is the local twist of the curve associated with the connection of the $U(1)$ gauge.

We consider low temperatures with only long wavelength spin waves being excited. Because the pulse represents a fast coherent rotation new spin waves with short wavelength are not exited during the pulse. Therefore the induced spin precession afterwards is classical long wavelength dynamics. The exchange part of the Hamiltonian of the ferromagnetic particle reads
\begin{equation}\label{ExchangeHamiltonian}
H_{ex}[\phi]=\frac{J}{2}\int\left((\partial_\mu\phi^a)^2+ \frac{(\phi_a\partial_\mu \phi^a)^2} {1-\phi_a\phi^a}\right)d^d\vec{r},
\end{equation}
where $J$ is the exchange constant and $d$ is the dimension of a ferromagnetic system. The Goldstone theorem forbids contribution to the exchange Hamiltonian from the uniform mode $\vec{n}$. Relativistic spin-orbit and dipole-dipole interactions induce the anisotropy energy that for simplicity we write in the local approximation:
\begin{equation}\label{AnisotropyHamiltonian}
H_{an}[\vec{M}]=\int E\left(\vec{M}(t,\vec{r})\right) d^d\vec{r},
\end{equation}
where the anisotropy density $E(\vec{m})$ is an arbitrary function of $\vec{m}$ (here and below $\vec{m}$ is an arbitrary unit vector distinct from the special time dependent $\vec{n}(t)$) and does not depend on the coordinate $\vec{r}$. One-ion magnetic anisotropy is an even function $E(\vec{m})=E(-\vec{m})$ and is symmetric under the crystallographic group of transformations $\hat{T}_i$: $E(\hat{T}_i\vec{m})= E(\vec{m})$. But a uniform magnetic field could be included also. The choice of anisotropy as a uniform function of local ferromagnetic order Eq.(\ref{AnisotropyHamiltonian}) is very general and applies to many experiments. It holds for particles of special shapes like the ellipsoid where the dipole-dipole interaction is described by effective uniform demagnetizing field. Also on the interfaces Neel predicted a quite general phenomenon of the interface anisotropy energy due to the break down of the translational symmetry in the direction perpendicular to the interface. In this case Eq.(\ref{AnisotropyHamiltonian}) holds  provided the spin wave modes with wave vectors perpendicular to the interface are not excited as for ferromagnetic thin films.

The dynamics of a ferromagnetic particle is described by the classical Larmor precession equation for local ferromagnetic order $\vec{M}(t,\vec{r})$:
\begin{equation}\label{Larmor}
\partial_t\vec{M}=\gamma\vec{M}\times\frac{\delta H}{\delta \vec{M}},
\end{equation}
where $\gamma=e/mc$ is the gyromagnetic ratio and $H=H_{ex}[\phi]+H_{an}[\vec{M}]$ is the Hamiltonian. The Larmor equation conserves the total energy $dH/dt=0$. We substitute into Eq.(\ref{Larmor}) $\delta H_{ex}/\delta \vec{M}= \vec{e}_a \delta H_{ex}/\delta \phi^a$ and $\delta H_{an}/\delta \vec{M}=-\vec{F}(\vec{M})$, where the force $\vec{F}(\vec{m})=-dE/d\vec{m}$ is a vector field on a sphere $\vec{m}^2=1$. In the Frenet-Serret gauge for the rotating frame: $\vec{e}_2\cdot\vec{F}=k(t)$ and $\vec{e}_1\cdot \vec{F}=0$. The total energy is the sum of the uniform mode energy $E(\vec{n})$ and the spin wave energy $\mathcal{E}_{sw}(\vec{n},[\phi])$: $H=E(\vec{n})+\mathcal{E}_{sw}(\vec{n},[\phi])$.

We project the Larmor equation Eq.(\ref{Larmor}) onto an equation representing the uniform mode Eq.(\ref{LLE}) and onto an equation representing the remaining non-uniform modes Eq.(\ref{SpinWaveEquation}). The component of both uniform and non-uniform equations that is parallel to $\vec{n}(t)$ is a linear combination of the other equations and therefore is redundant. We assume that the spin wave modes are weakly excited and expand the Larmor equation up to the second power of spin wave amplitudes. We find the equation for the uniform mode:
\begin{equation}\label{LLE}
\frac{d\vec{n}}{dt}=\vec{n}\times \left(-\vec{F}- \vec{n} \times \vec{F}\ \frac{1}{k^2} \frac{d}{dt}\mathcal{E}_{sw}\right).
\end{equation}
Here $\mathcal{E}_{sw}(\vec{n},[\phi])$ is the spin wave energy in the harmonic approximation:
\begin{equation}\label{SWEnergy}
\mathcal{E}_{sw}=\frac{1}{2}\int\left(J(\partial_\mu\phi^a)^2 +\phi^a C_{ab}\phi^b -\vec{n}\cdot \vec{F}\phi_a\phi^a \right) d^d\vec{r},
\end{equation}
where the anisotropy energy mass tensor is:
\begin{equation}\label{MassTen}
C_{ab}(\vec{n})=e_a^\alpha\frac{\partial^2 E}{\partial n^\alpha\partial n^\beta} e_b^\beta .
\end{equation}
The gauge dependent component of the force $\vec{n}\cdot\vec{F}$ and $C_{11}$ are related to the precession twist $\kappa(t)$. In the limit of zero spin wave amplitudes the conservation of energy $E(\vec{n})$ in the rotating Frenet-Serret frame gives: $\vec{n}\cdot\vec{F}- C_{11}\left(\vec{n}\right)=\kappa(t)$. $C_{22}\left(\vec{n}\right)- C_{11}\left(\vec{n}\right)$ and $C_{12}\left(\vec{n}\right)$ are both gauge independent. Obviously, the uniform mode Eq.(\ref{LLE}) has an integral: $E(\vec{n})+ \mathcal{E}_{sw}(\vec{n},[\phi])= const$, which is the total energy conservation.

Note that Eq.(\ref{LLE}) has the form of the Landau Lifshitz equation with the Gilbert dissipation constant
\begin{equation}\label{GilbertDamping}
G(t)= \frac{1}{k^2(t)} \frac{d}{dt}\mathcal{E}_{sw}(\vec{n},[\phi])
\end{equation}
The sign of the Gilbert constant can be established in the vicinity of the maximum or minimum point of the anisotropy energy $E(\vec{n})$  under the condition of growing amplitudes of spin waves. In the vicinity of the minimum point both the curvature $k(t)>0$ and the spin-wave mass tensor $C_{ab}(t)$ approach zero and are proportional to the angle of the deviation from the minimum point. On the contrary the twist $\kappa(t)$ approaches some finite negative value. In Eq.(\ref{SWEnergy}) we retain only the leading large term $\mathcal{E}_{sw}= -\kappa(t)\int \phi_a(t,\vec{r}) \phi^a(t,\vec{r}) d^d\vec{r}$. For growing spin wave amplitudes and negative $\kappa$: we find $d\mathcal{E}_{sw}/dt>0$. The Gilbert constant is positive and due to the condition $E(\vec{n})+\mathcal{E}_{sw}=const$: $dE(\vec{n})/dt<0$. Therefore the vector $\vec{n}$ approaches the minimum point during the precession. Superficially this could be seen as the dissipation process although the non-uniform Larmor precession in the ferromagnetic system conserves energy and is fully time-reversible. The same analysis shows that near the maximum point the twist $\kappa(t)$ approaches some finite positive value and therefore the Gilbert constant is negative, $dE(\vec{n})/dt>0$ and the vector $\vec{n}$ approaches the maximum point during the precession that is accompanied by the spin-wave instability.

We find also the equation for the non-uniform mode:
\begin{equation}\label{SpinWaveEquation}
\epsilon_{ab}\partial_t\phi^b=-J\partial_\mu^2\phi_a+C_{ab}(\vec{n})\phi^b+\left(\kappa -\vec{n}\cdot\vec{F}\right) \phi^a,
\end{equation}
or in the principal normal and binormal coordinates:
\begin{eqnarray}\label{SWE}
\partial_t\phi_1 &=& -J\partial_\mu^2\phi_2+\left(C_{22}(\vec{n})-C_{11}(\vec{n})\right)\phi_2+C_{21}(\vec{n})\phi_1 \nonumber\\
-\partial_t\phi_2 &=& -J\partial_\mu^2\phi_1+C_{12}(\vec{n})\phi_2.
\end{eqnarray}
This system is linear and Hamiltonian. The coefficient $C_{12}(\vec{n})$ can be related to the curvature $k(t)$. We note that $k\vec{e}_1= [\partial E/\partial\vec{n}\times \vec{n}]= d\vec{n}/dt$ and $k\vec{e}_2=\partial E/\partial \vec{n}- \vec{n}(\vec{n}\cdot \partial E/\partial \vec{n})$, and therefore 
\begin{equation}\label{C12}
C_{12}(\vec{n})= \frac{1}{2k(\vec{n})^2}\left(\frac{d\vec{n}}{dt}\cdot\frac{d}{d\vec{n}}\right) k^2(\vec{n})=\frac{d}{dt}\log{k(t)}.
\end{equation}
In the limit of infinitely small amplitudes of all spin waves the anisotropy energy $E(\vec{n})$ is approximately conserved and the uniform mode precesses periodically in time $\vec{n}(t)= \vec{n}(t+T)$, where $T$ is the period of precession. Solution of the Eqs.(\ref{SWE}) for a spin wave mode with momentum $\vec{q}$ can be formally written as a monodromy matrix:
\begin{equation}\label{monodr}
 \vec{\phi}(t+T)=\hat{\mathcal{M}}(t,T,\vec{q})\vec{\phi}(t)
\end{equation} 
where $\vec{\phi}(t)=(\phi_1(t),\phi_2(t))$. Monodromy matrix has unit determinant $\det{\mathcal{M}(t,T,\vec{q})}=1$. Let $a=\textrm{Tr} \mathcal{M}/2$. If $|a|< 1$ then the eigenvalues of $\mathcal{M}$ are two complex numbers $\lambda_{1,2}=\exp(\pm i\arccos(a))$ otherwise ($|a|\geq 1$) they are real: $\lambda_{1,2}=\exp(\pm \textrm{arccosh}(|a|))$. In the later case spin wave mode is instable and its amplitude grows exponentially with the increment $\nu(\vec{q})= \log|\lambda_1|$. Two examples have been studied numerically: a) $E(\vec{m})=\lambda_x m_x^2+\lambda_y m_y^2+\lambda_z m_z^2$ and b) $E(\vec{m})=m_x^4+m_y^4+m_z^4$. In the case a) the uniform precession: $\vec{n}(t)$, $\vec{e}_1(t)$, $\vec{e}_2(t)$ could be found analytically in terms of the elliptic functions. Numerical solution in both cases a) and b) shows that $a(\lambda)$ is oscillating (but not periodic) function of spectral parameter $\lambda=J\vec{q}^2$. Note, that for $C_{ab}=0$, $a(\lambda)=\cos(\lambda T)$. There are infinitely many intervals of instability (zones $|a(\lambda)|\geq 1$) in both cases, they appear periodically and for $\lambda\rightarrow\infty$ they shrink. We propose that for any initial condition precession in arbitrary $E(\vec{m})$ there exists unstable spin-wave modes in momentum space.

In the extremum point the uniform mode $\vec{n}$ is constant and the curvature tensor $C_{ab}$ does not depend on time. In the adiabatic limit of very slow precession in the vicinity of extremum point the spin wave dispersion reads:
\begin{equation}\label{SWdispersion}
\omega(\vec{q})=\sqrt{Jq^2(Jq^2+C_{22}-C_{11})-C_{12}^2}.
\end{equation} 
The spin wave instability develops if either a) $C_{12}\neq 0$ or b) $C_{22}-C_{11}<0$. According to the condition b) systems with an easy-axis anisotropy always instable whereas easy-plane systems could be stable provided $C_{12}=0$ that according to the Eq.(\ref{C12}) requires a precession along a circle with the constant curvature. Thus a paradoxical situation arises. Let excite a however small angle precession of easy-axis ferromagnet around the equilibrium point at such low temperature that amplitudes of thermal spin waves are much less than the angle of precession. According to Eq.(\ref{SWdispersion}) and according to our numerical simulation of very long (2048 sites) one dimensional ferromagnet chain a spin-wave instability will develop. But spin wave dispersion in the laboratory (non-rotating) frame is $\omega(\vec{q})= Jq^2+\lambda$, where positive $\lambda$ is the strength of easy-axis anisotropy and is the gap of spin wave excitation that naively could not be bridged by a small deviation from the equilibrium  point.  

The spin wave dispersion (\ref{SWdispersion}) has a branch with positive imaginary part: $\omega_A(\vec{q})=i\nu(\vec{q})$ at the region of wave vectors $0<|\vec{q}|<q_+$, where $2Jq^2_+= \sqrt{(C_{22}-C_{11})^2+4C_{12}^2} -(C_{22}-C_{11})$ and the index $A$ stands for advanced. The spin-wave field $\phi_a(t,\vec{q})$ grows exponentially with the increment $\nu(\vec{q})= \textrm{Im}\, \omega_A(\vec{q})$ reaching its maximum at $|\vec{q}|=q_m$ where in the case b) $2Jq^2_m=C_{11}-C_{22}$ and in the case not b) $q_m=0$. The spin wave distribution function in momentum space: $N_{\vec{q}}(t)= \phi_a(t,\vec{q}) \phi^a(t,\vec{q})$, grows accordingly. The spin wave energy Eq.(\ref{SWEnergy}) is the sum over all modes: $\mathcal{E}_{sw}(t)= \sum_{\vec{q}} \mathcal{E}_{sw}(t,\vec{q})$, where using Eq.(\ref{SWE}) we find the energy of one spin wave mode: $\mathcal{E}_{sw}(t,\vec{q})=-\kappa N_{\vec{q}}(t)$.

Growing instability disturbs the distribution of spin waves in the momentum space and makes it different from the thermal equilibrium distribution in momentum space $N^0_{\vec{q}}= T/\omega_0(\vec{q})$, where the real and positive $\omega_0(\vec{k})$ is the spin wave dispersion in the stable state before the pulse has arrived. After the pulse the spin wave instability is developing and it in turns gives rise to the spin wave scattering in the direction that restore the thermal distribution and both these processes could be described by the Boltzmann kinetic equation:
\begin{equation}\label{BoltzmannKE}
\partial_t N_{\vec{q}}= 2\nu(\vec{q}) N_{\vec{q}} -I(N_{\vec{q}}),
\end{equation}
where $\nu(\vec{q})=\textrm{Im}\,\omega_A(\vec{q})$ and the relaxation rate is given by the Boltzmann collision integral $I(N_{\vec{q}})$. In our situation obviously the rate of instability and the collision integral in the Boltzmann equation are of the same magnitude. The leading contribution to the collision integral comes from the two magnon scattering with one magnon having large momentum $\vec{p}$ in the incoming and outgoing states and second long wavelength magnon with the momentum $\vec{q}$ in the instability domain of phase space:
\begin{equation}\label{CollisionIntegral}
I(N_{\vec{q}})=\int W(\vec{q},\vec{q'})(N(\vec{q})-N(\vec{q}))\frac{d^d\vec{q}}{(2\pi)^d},
\end{equation}
where the scattering probability does not depend on momenta: $W(\vec{q},\vec{q'}) =W$, and it is  is small at low temperatures $W\sim (T/J)^{5/2}$. It could be proven that the probability of absorption of long wavelength magnon with purely imaginary dispersion $\omega_A(\vec{q})=i\nu(\vec{q})$ by a short wavelength magnon as well as the probability of emission of such a magnon in the process of two short wavelength magnons collision are both zero. The probability of two long wavelength magnon scattering is proportional to the small factor $(C/J)^2$ and could be neglected.

\begin{figure}[t] \label{figure}
\includegraphics{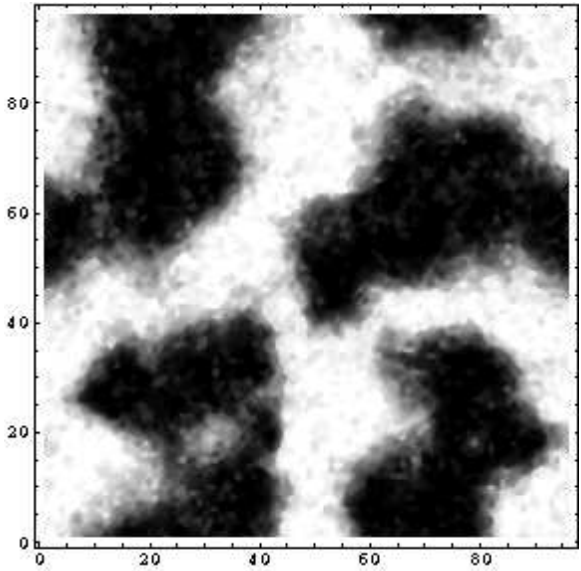}
\caption{A snapshot of chaotic time dependent domains in the developed stage of domain instability in the easy axis: $E(\vec{m})=  -\lambda_z m_z^2$, cluster of 96x96 exchange coupled classical spins. Non-linearity $\vec{M}^2(\vec{r},t)=1$ caps the growing amplitude of the spin waves into a domain picture. Shades of gray represent $z$-component magnetization with black and white being spin up and down.} 
\end{figure}

Solution of the Boltzmann kinetic Eq.(\ref{BoltzmannKE}) yields:
\begin{equation}\label{BKESolution}
N_{\vec{q}}(t)=\exp\left((2\nu(\vec{q})-\chi)t\right)\left(N_{0\vec{q}}+\Phi(t)\right),
\end{equation}
where $\Phi(t)=N_0(\exp(\chi t)-1)$, $\chi=W\Omega$ is the relaxation rate, $\Omega=\int_{q<q_+} d^d\vec{q}/(2\pi)^d$ is the volume of instability region in the momentum space and $N_0=\int_{q<q_+} N_{0\vec{q}}\, d^d\vec{q}/(2\pi)^d/ \Omega$ is the average number of would be instable spin waves in the system per one mode before the pulse arrives. Solution of Eq.(\ref{BKESolution}) has an asymptote: $N_{\vec{q}}(t)= N_0\exp(2\nu(\vec{q})t)$, as $t\rightarrow\infty$ that does not depend on the initial occupation number for this mode. The spin wave energy Eq.(\ref{SWEnergy}) with $\phi_a(t)$ from Eq.(\ref{SWE}) reads:
\begin{equation}\label{SWEnergyInst}
\mathcal{E}_{sw}(t)=-V\kappa \int_{q<q_+} N_{\vec{q}}(t) \frac{d^d\vec{q}}{(2\pi)^d},
\end{equation}
where $V$ is the volume of the particle and $\kappa$ is the twist of precession curve. For large $t$ the spin wave energy could be estimated by the steepest descent method.

Small size of a particle suppresses the domain instability growth. The maximum wavevector of the instability region could be shorter then the smallest size-quantized wavevector in finite particle. The critical size $L_c$ when domain instability could develop depends on the shape of the particle and is proportional to the domain wall width $L_c\sim\sqrt{J/\lambda}$, where $\lambda$ is the typical anisotropy energy. We studied numerically the case of of 2D square particle with periodical boundary condition (torus) with the easy-axis perpendicular anisotropy: $E(\vec{m})=-\lambda m_z^2$. For initial state $n^z=0$, we find the critical size $L_c\approx 7\sqrt{J/\lambda}$. 

In conclusion we have found the microscopic equations that govern spin wave instability in the non-equilibrium precessional state of a ferromagnetic system. In the developed stages of instability spin-waves saturate to a time dependent chaotic domain pattern. This multidomain state Fig.1 lives until the intrinsic relaxation of the ferromagnet eventually establishes a single domain ground state, randomly if the ground state is degenerate. This helps to understand stochasticity of precessional switching observed in Ref.\cite{tudosa:04}.

I am greatful to Stanford Linear Accelerator Center for hospitality and to H.C. Siegmann and J. Stohr for interest to this work, discussions and suggestions.

\end{document}